\documentclass[runningheads,citeauthoryear]{apinv}
\usepackage{epsfig,cite,graphics}
\usepackage[T2A]{fontenc}
\usepackage[cp1251]{inputenc}

\usepackage{graphicx}

\def\lsi{LSI+61$^{\circ}$303}
\def\4u{4U~2206+54} 
  
\def\kms{km~s$^{-1}$}  

\def\vsi{$v\: \sin i$}  
\def\rsun{R$_{\odot}$} 
 
\def\msun{M$_{\odot}$} 
\def\Ha{H$\alpha$}

\begin{document}

\title{Circumstellar discs in X/$\gamma$-ray binaries: first results from the Echelle spectrograph}
\titlerunning{Echelle spectra of X/$\gamma$-ray  binaries}
\author{R. Zamanov\inst{1}, K. Stoyanov\inst{1}, J. Mart{\'{\i}\inst{2}}}
\authorrunning{Zamanov, Stoyanov, Mart{\'{\i}}}
\tocauthor{R. Zamanov, K. Stoyanov, J. Mart{\'{\i}}}
\institute{Institute of Astronomy and National Astronomical Observatory, Bulgarian Academy of Sciences, Tsarighradsko Shose 72, 
        BG-1784 Sofia, Bulgaria
	\and Departamento de F\'isica (EPSJ), Universidad de Ja\'en, Campus Las Lagunillas,  A3-420, 23071, Ja\'en, Spain
        \newline
	\email{rkz@astro.bas.bg, kstoyanov@astro.bas.bg, jmarti@ujaen.es}    }
\papertype{Submitted on 12.08.2015; Accepted on 08.09.2015}	
\maketitle

\begin{abstract} 
Here we report our first spectral observations of  Be/X-ray and $\gamma$-ray binaries obtained with the new Echelle spectrograph 
of the National Astronomical Observatory Rozhen. For four objects (LSI+61$^{\circ}$303,
$\gamma$~Cas, MWC~148, 4U~2206+54), 
we report the  parameters and estimate the sizes of their circumstellar discs using
different emission lines (H$\alpha$, H$\beta$, H$\gamma$, HeI and FeII).
For MWC~148, we find that the compact object goes deeply through the disc. 
The flank inflections of H$\alpha$ can be connected with inner ring formed at the periastron passage 
or radiation transfer effects.    
We point out an intriguing similarity between  the optical emission lines  of the 
$\gamma$-ray binary MWC~148 and the well known Be star $\gamma$ Cas. 
\end{abstract}
\vskip 0.1cm 
\keywords{ Stars: emission-line, Be -- Stars: winds, outflows -- X-rays: binaries -- Stars: individual: LSI+61303,
$\gamma$ Cas, MWC~148, 4U~2206+54 }

\section{Introduction}

The Be-X/$\gamma$-ray binaries (BeX$\gamma$B) are systems that consist of a compact object orbiting
an optical companion that is an OBe star. Oe and Be stars are non-supergiant fast-rotating O-type and B-type and 
luminosity class III-V stars which, at some point of their lives, have shown spectral lines in emission
(Porter \& Rivinius 2003; Balona 2000; Slettebak 1988). The best studied lines are those of hydrogen (Balmer and Paschen series),
but  the OBe stars may also show He and Fe in emission (see Hanuschik 1996, and references therein). 
They also typically display an amount of  infrared excess. The origin of the emission lines and infrared excesses in BeX$\gamma$B
is attributed to an equatorial disc, fed from material expelled from the rapidly rotating OBe star. During periastron, 
the compact object passes close to this disc, and sometimes may even go through it causing a major disruption. 
A large flow of matter is then captured by  the compact object. In  most cases the compact object is a neutron star 
detected as an X-ray pulsar (see Bildsten et al. 1997, and references therein). The conversion of the kinetic energy 
of the infalling matter into radiation powers the X-rays. Particle acceleration processes associated with the interaction
between the compact object and the circumstellar disc can also take place, rendering the system as a High-Energy (HE)
and even Very-High Energy (VHE) source.

In this paper, we present the current status of a long-term project aimed to monitor the variability of emission lines
in a selected sample of OBe binaries using high resolution spectroscopy. 
Two of the targets reported here, MWC 148 and \lsi, belong to the 
$\gamma$-ray binary family.  Only a handful of these objects are currently known where the bulk of their
luminosity is radiated beyond 1 MeV (see e.g. Dubus 2013). The spectral properties of optical emission lines, specially in the H$\alpha$ region, are good
tracers of the physical structure of the circumstellar discs around the optical companion. It is expected that 
our observations will enable to establish the expected correlation between
the spectral line properties and the HE/VHE emission.

%
%

\section{Observations and Objects}
The data reported here are from the new EsPeRo spectrograph.  
EsPeRo  is fibre-fed Echelle spectrograph attached to the 2.0 m  Ritchey-Chr\'etien-Coud\'e  (RCC) telescope of the National Astronomical Observatory Rozhen (NAO). 
The spectra are reduced in the standard way including bias removal, flat-field correction, 
wavelength calibration and correction for the Earth's orbital motion. 
Pre-processing of data and parameter measurements are performed using various routines provided by the 
IRAF
software package.
The wavelength calibration is performed using a  Thorium-Argon (ThAr) lamp. 
The light source is a  1.5" diameter  PHOTRON Hollow Cathode Lamp.
 To facilitate future data reduction, a few Echelle orders with identified prominent features of  the ThAr lamp
are given in the Appendix. At 6560~\AA,
the spectrograph provides a  resolution of  0.056~\AA/pixel, with the Full Width Half Maximum (FWHM) of the 
ThAr lines being  0.22~\AA. Similarly,  
at 4800 \AA, we obtain a spectral resolution of 0.041~\AA/pixel with a  0.16~\AA\ FWHM of the calibration lamp lines.

\begin{table*}[htb]
  \begin{center}
  \caption{Journal of observations. In the table are given as follows:  object name, 
  V band magnitude, spectral type,  start of the observations, exposure time, signal-to-noise ratio at 6600~\AA,
  and system orbital period.  }
  \begin{tabular}{p{2cm} p{1cm}  p{1.5cm}   p{3cm}   p{2.0cm} p{1.5cm} p{2cm} cccccccc}
  Name         &  V	&   Spec.     & Date-Obs. 	  & Exp-time &  S/N  &   $P_{orb} [d]$  &    \\
  \\
  $\gamma$~Cas &   2.4  &  B0.5 IVpe & 2014-01-14 00:43   &   15 min &   90  &   203.59	 &    \\
  MWC~148      &   9.1  &  B0 Vpe    & 2014-03-13 20:03   &   60 min &   65  &   321		 &    \\
  \lsi         &  10.8  &  B0 Ve     & 2014-02-17 19:23   &   60 min &   23   &   26.496	 &    \\
  \4u          &   9.9  &  O9.5 Vep  & 2014-01-14 16:12   &   60  min &   22   &   9.568 	 &    \\
  \end{tabular}
  \label{table1}
  \end{center}
\end{table*} 

The list of observations is given in Table~\ref{table1}. The resulting, normalized spectra are presented
in Fig. \ref{fig.e} for different emission lines of interest. 
The measured equivalent widths and the distance between the peaks are summarized in Table~\ref{table2}. 
Here we provide a short description of the 
four binary stars observed at the beginning of 2014.

\subsection*{$\gamma$~Cas }
$\gamma$~Cas  is a well known,  bright ($V \sim 2.3$~mag, B0.5IVpe) Be star. 
It is the first emission line star discovered by one of the founders of the study of stellar spectra, Father Angelo Secchi
(Secchi 1866).
The orbital period is 203.52~d and the orbit seems to be circular (Miroshnichenko, Bjorkman \& Krugov 2002, Nemravov{\'a} et al. 2012). 
Harmanec et al. (2000) reported the mass of the donor being  $\approx 13$~\msun\, and the
inclination angle  $\it i \approx 45^0$.
The orbital motion is  detectable on spectra and 
implies that the Be star is the primary component of a spectroscopic binary system.
The compact object has a mass of about 1~\msun, appropriate for a white dwarf or a neutron star,
but it could also be a normal late-type dwarf (Harmanec et al.  2000; 
Nemravov{\'a} et al. 2012, Miroshnichenko et al. 2002), or a neutron star 
in the propeller regime (Shrader et al. 2015),  or a black hole (low mass, slowly rotating, non-magnetized).  
Robinson \& Smith (2000) found that the X-ray flux varied with a period P = 1.12277 d, which they
interpreted as the rotational period of the mass donor.
Here we assume $R_1 = 8.7 \pm 0.7 $~\rsun\  and \vsi$= 385$~\kms\  (Harmanec 2002) for the radius and projected rotational velocity
of the donor star, respectively.

\subsection*{MWC~148}
MWC~148  is a B0Vpe star which is the optical counterpart of the TeV (tera-electron volt) source HESS J0632+057
(Aharonian et al. 2007). The identification of  $\gamma$-ray source with the optical Be star 
is a result of the detection of radio (Skilton et al. 2009) and X-ray emission associated 
with the object (Hinton et al. 2009; Falcone et al. 2010) with similar spectral indexes and variability as observed
in $\gamma-$ray binaries PSR~B1259-63, LS~5039 and \lsi\ (Kargaltsev et al. 2014). Extended and variable radio emission at AU scales, displaying 
similar morphology as in these three X-ray binaries, has been reported  (Mold{\'o}n, Rib{\'o} \& Paredes 2011).
The absence of significant radial velocity shifts in the B0 star supports a long period $>100$ d (Aragona, McSwain, \& De Becker 2010). 
Bongiorno et al. (2011) report strong X-ray flares in Swift/XRT data, 
modulated with  $321 \pm 5$~d, which most probably is the binary period. Casares et al. (2012) estimated the radius of the primary 
$R_1 = 6 - 10$~\rsun\  and \vsi$= 373$~\kms.  

\subsection*{\lsi}
LSI$+61{^0}303$ consists of  a massive  B0Ve star and a compact object. The orbital period is $26.4917 \pm 0.0025$~d, obtained 
with Bayesian analysis of radio data (Gregory, Peracaula \& Taylor 1999; Gregory 2002).
The nature of the compact object  remains a mystery even after 37 years of observations over a 
wide range of  wavelengths.  
Most probably it is a  neutron star, but it might  be a  magnetized  black hole (Punsly  1999) 
acting as a precessing microblazar (Massi, Ros \& Zimmermann 2012).  
Here we assume a primary radius value of  $R_1 = 6.7$~\rsun\  (Grundstrom et al. 2007) and 
\vsi$= 349 \pm 6$~\kms\ (Zamanov et al. 2013).

\subsection*{4U~2206+54 }
4U 2206+54 is a persistent high-mass X-ray binary star at a distance of 2.6 kpc (Blay  et al. 2006). 
It was first detected as an X-ray source by the $Uhuru$ satellite (Giacconi et al. 1972). The mass donor is classified as an O9.5Vp star with a higher
than normal helium abundance, underfilling its Roche lobe and losing mass via a slow but dense stellar wind, flowing at
350 \kms\ (Rib{\'o} et al. 2006). However, there are some metallic
lines typical for a later-type spectrum (Negueruela \& Reig 2001). 
The compact object is a neutron star (Torrej{\'o}n et al. 2004) with spin period $P_{spin} = 5554 \pm 9$~s (Finger et al. 2010).
The X-ray spectrum of 4U 2206+54 is typical for a neutron
star accreting material onto its magnetic poles. The orbital period of the system is $P_{orb} = 9.5591 \pm 0.0007 $~d
(Rib{\'o} et al. 2006)  and  the orbit is elliptical with $e = 0.30$ (Stoyanov et al. 2014). 
For this object we adopt the values $R_1 = 7.3$~\rsun\  (Rib{\'o} et al. 2006) and 
\vsi$= 315 \pm 70$~\kms\ (Blay  et al. 2006).

\begin{table*}[htb]
  \begin{center}
  \caption{Measured emission lines parameters. }
  \begin{tabular}{llrrrrrrrrcccccccc}
\hline
object     &  line               & $\; \;$ &  EW    & $\;$ & $\Delta V$ &    $\, \,$  & $R_{disc}/ R_1$  &  $R_{disc}$ & \\
           &                     &      &  [\AA] &  & [\AA]      &	     &  		&   [\rsun]   & \\
  \hline
     \\
$\gamma$~Cas & H$\alpha$         &    &  25.8	&   & 1.3?  & &  167?	 &     1400?  & \\
             & H$\beta$          &    &  3.77	&   & 2.16  & &   33	 &	290   & \\
             & H$\gamma$         &    &  0.86	&   & 2.60  & &   18.4   &	160   & \\
             & HeI$\lambda$6678  &    &  0.23	&   & 6.55  & &    6.8   &	 60   & \\
             & HeI$\lambda$5876  &    &  0.40	&   & 4.25  & &   12.5   &	110   & \\
             & FeII$\lambda$6317 &    &  0.39	&   & 4.47  & &   13.1   &	114   & \\
             & FeII$\lambda$5316 &    &  0.28	&   & 3.24  & &   17.7   &	154   & \\
             & FeII$\lambda$5169 &    &  0.62	&   & 3.26  & &   16.5   &	144   & \\
  \hline		     	     			      
  \\  
MWC~148      & H$\alpha$         &    &  34.4	&   & 2.46  & &  44.0	& 352  & \\
             & H$\beta$          &    &  4.44	&   & 2.38  & &  25.8	& 206  & \\
             & H$\gamma$         &    &  1.40	&   & 2.60  & &  17.2	& 137  & \\
             & HeI$\lambda$6678  &    &  0.27	&   & 6.24  & &   7.1	&  56  & \\
             & HeI$\lambda$5876  &    &  0.39	&   & 4.67  & &   9.8	&  78  & \\
             & FeII$\lambda$6317 &    &  0.48	&   & 4.43  & &  12.6	& 101  & \\
             & FeII$\lambda$5316 &    &  0.55	&   & 4.25  & &   9.7	&  77  & \\
             & FeII$\lambda$5169 &    &  1.10	&   & 3.63  & &  12.5	& 100  & \\
\hline	
\\				     												     
\lsi         & H$\alpha$         &    &  8.47	&   & 6.56  & &    5.42  &     36.3  &  \\
             & H$\beta$          &    &  0.67	&   & 6.62  & &    2.92  &     19.5  &  \\
\hline 
\\     				     													     
\4u          &  H$\alpha$        &    &  0.23	&   & 10.61 & &    1.69  &     12.3  & \\
 \\
\hline
  \end{tabular}
  \label{table2}
  \end{center}
\end{table*} 

 \begin{figure*}    
   \vspace{17.5cm}   
     \includegraphics{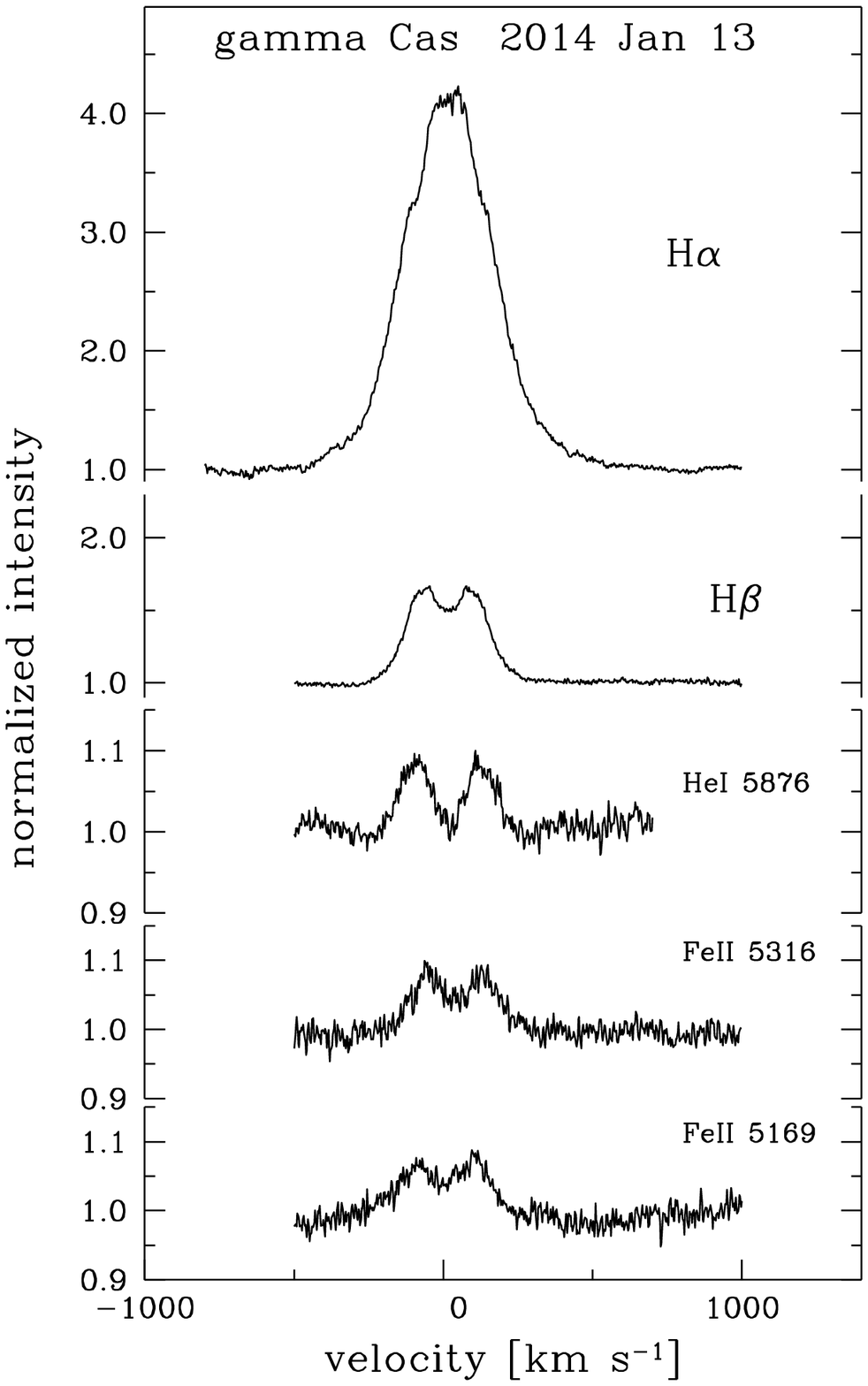}    
     \includegraphics{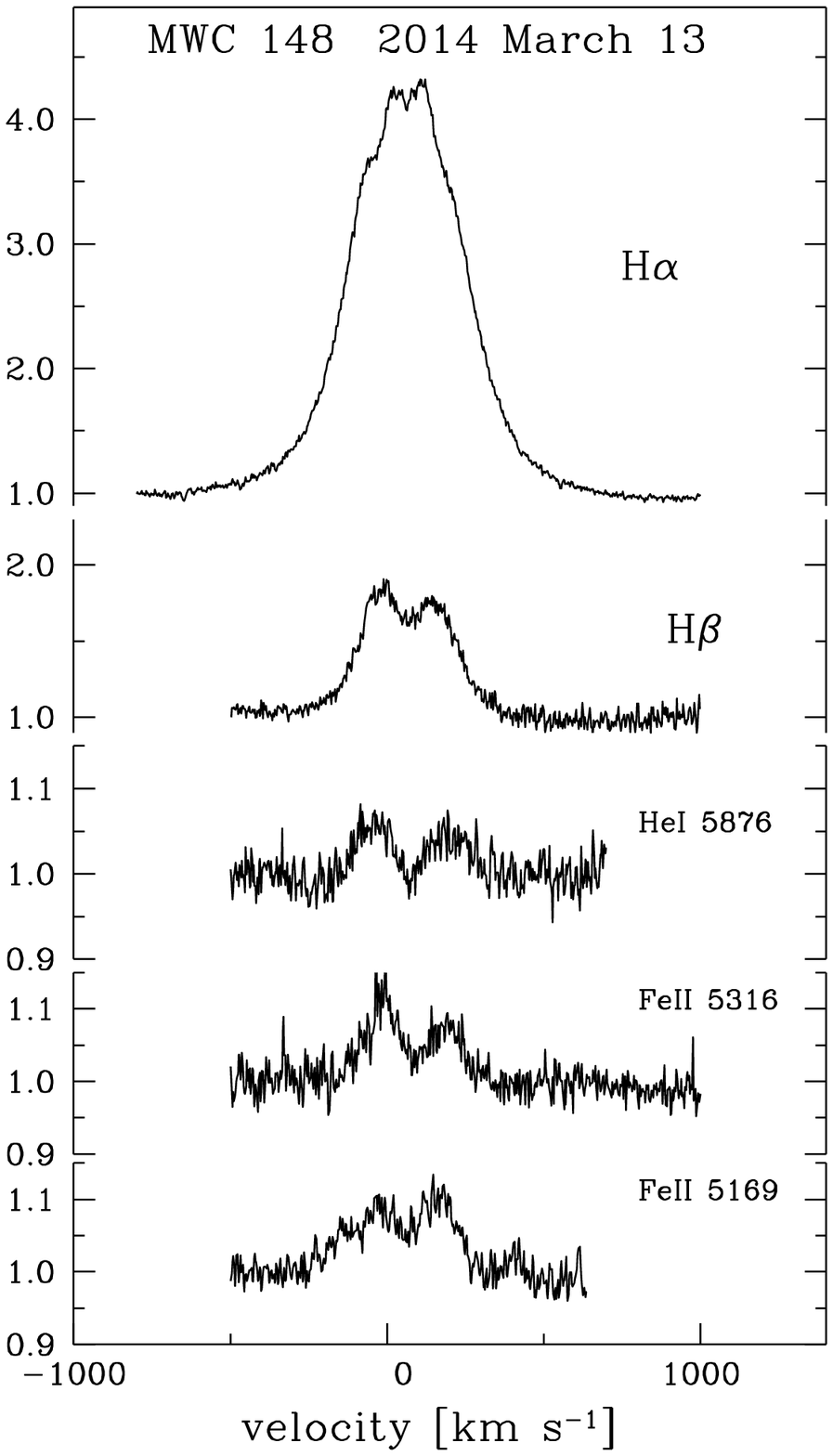}  
     \includegraphics{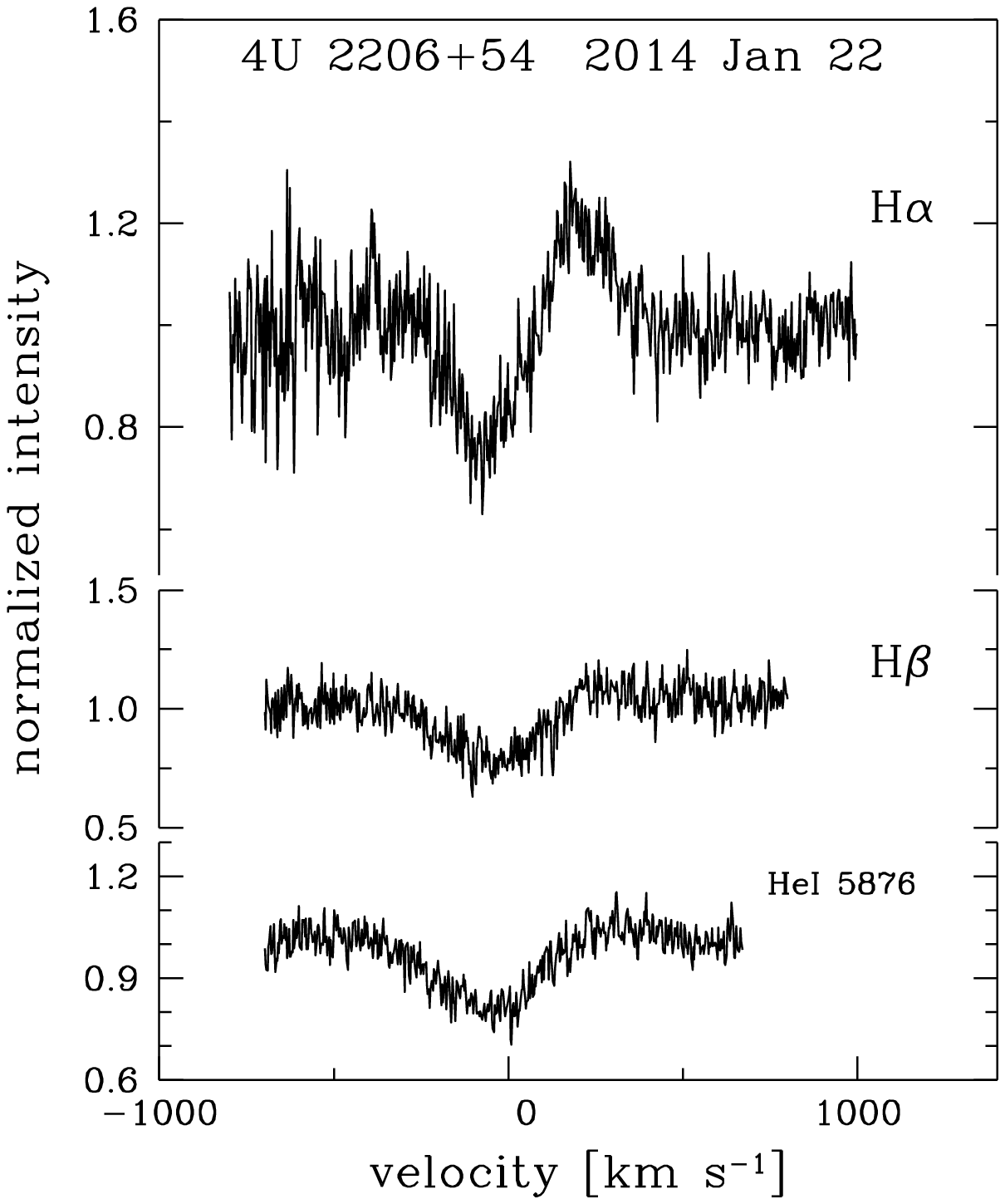}  
     \includegraphics{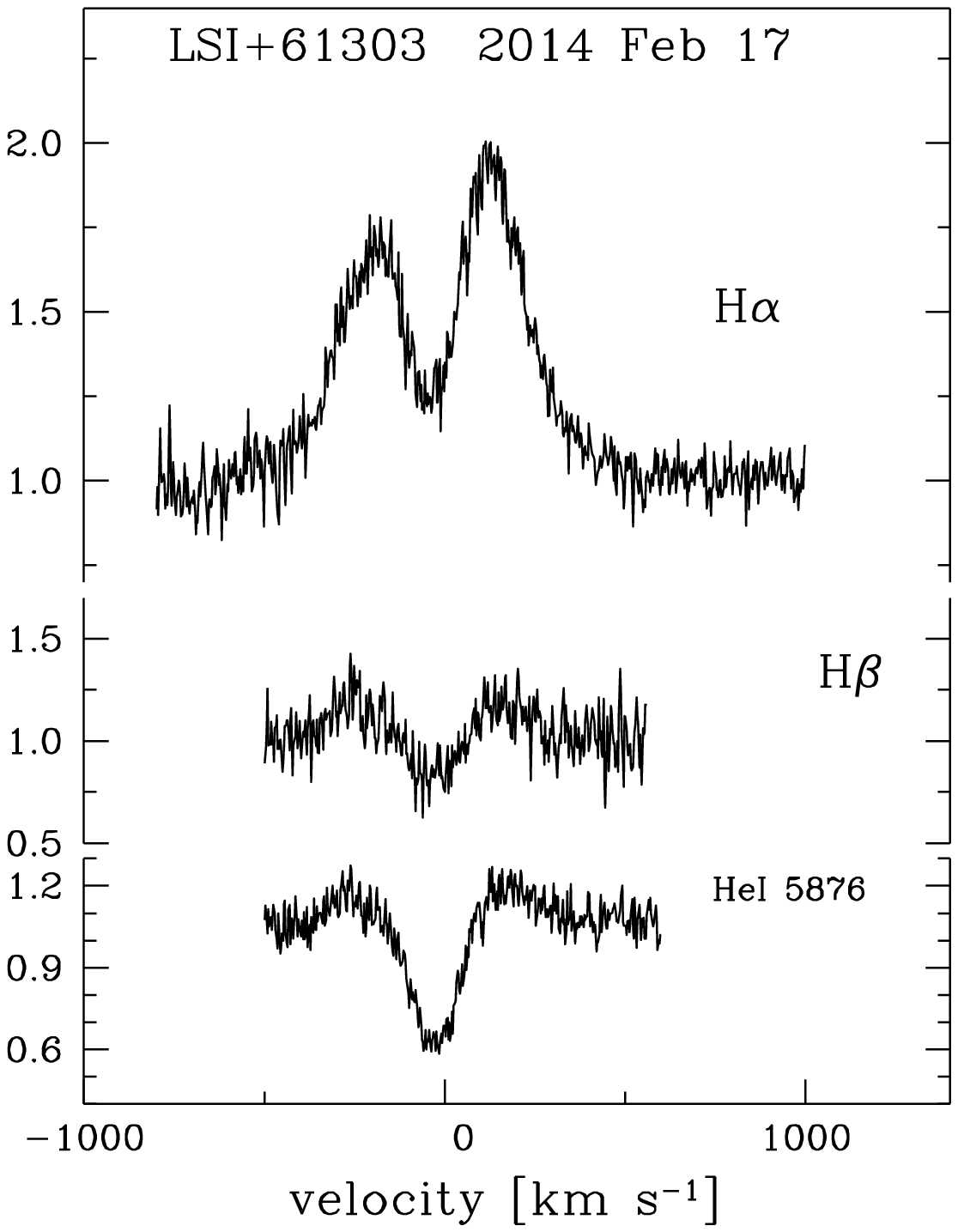}  
   \caption[]{ Emission line profiles (H$\alpha$, H$\beta$, HeI, FeII) of four high-mass X-ray binaries ($\gamma$ Cas, MWC~148, 4U~2206+54, and \lsi).
   Each spectrum is normalized to the local continuum.   It is remarkable that both
   $\gamma$~Cas and MWC~148 show wine-bottle profile in their respective H$\alpha$ lines.  A clear similarity 
   between the emission lines of these two objects  is also visible.
    The HeII 4686 \AA\ line is not detected in any of the four stars.
   }
   \label{fig.e}      
 \end{figure*}	     

 \begin{figure*}    
   \vspace{6.0cm}   
   \includegraphics{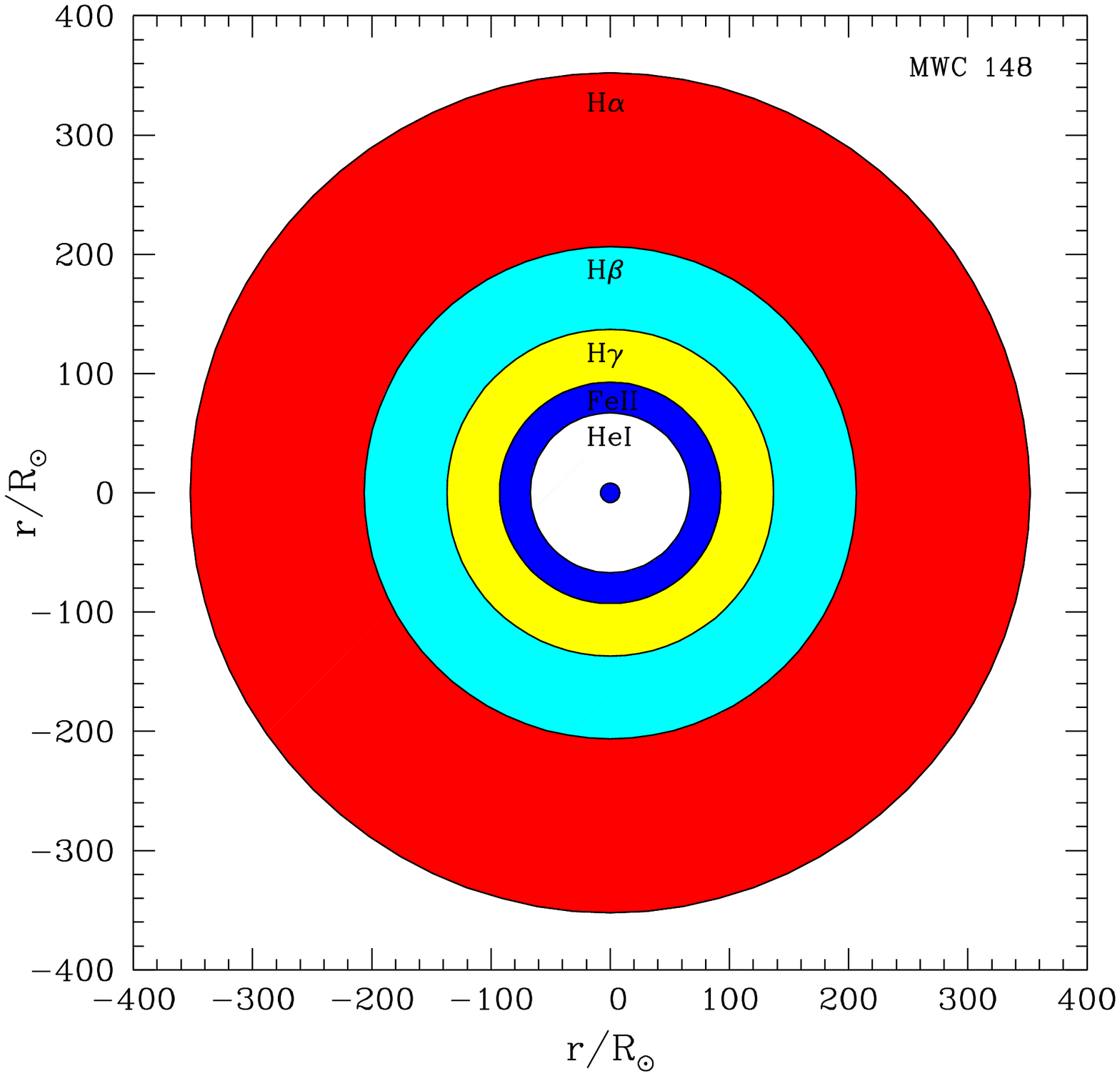}    
   \includegraphics{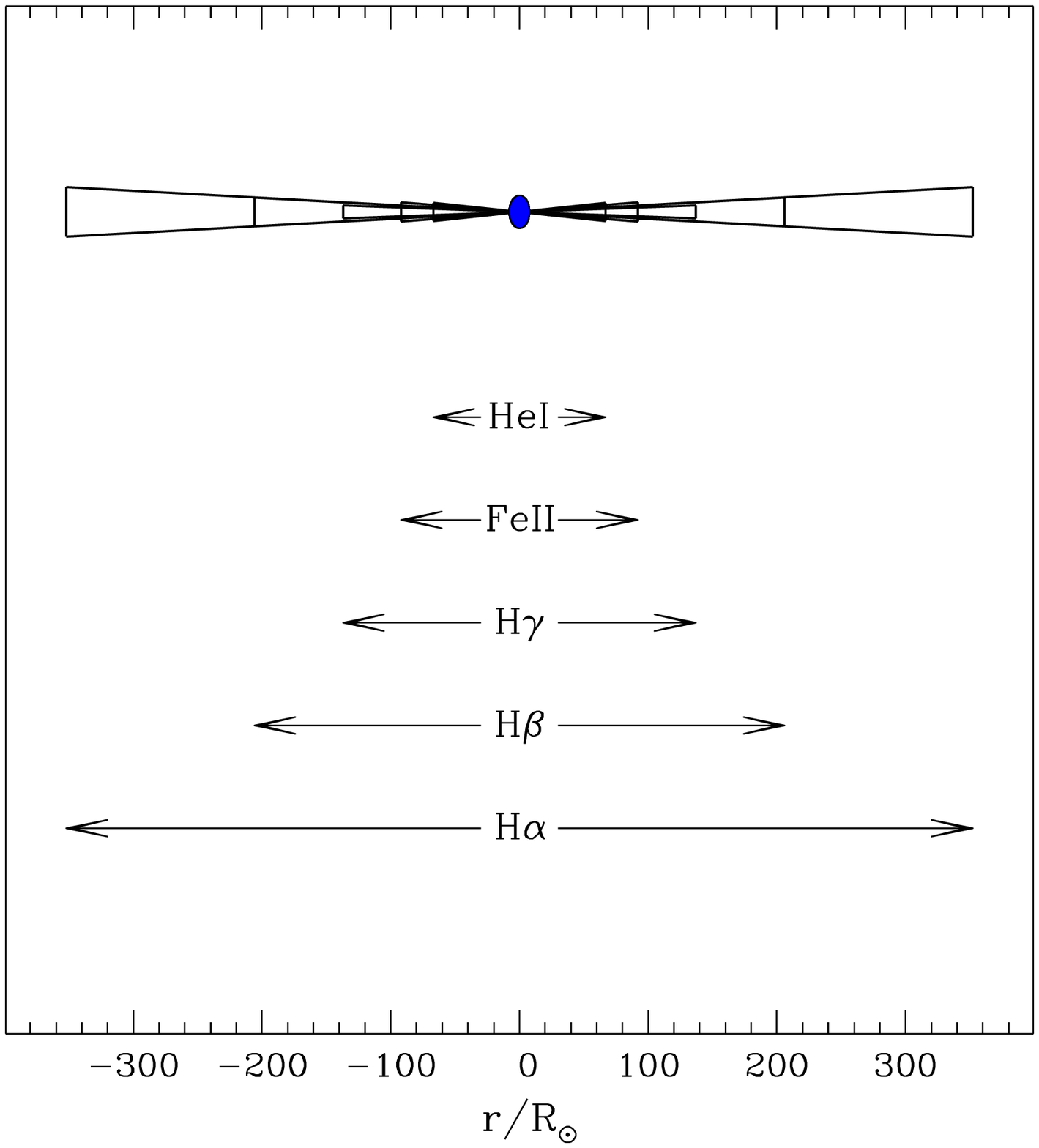}  
   \caption[]{ The sizes of the disc visible in different emission lines in MWC 148 seen both face-on (left)
    and edge-on (right).  }
   \label{fig.MWC148}      
   \vspace{7.0cm}   
   \includegraphics{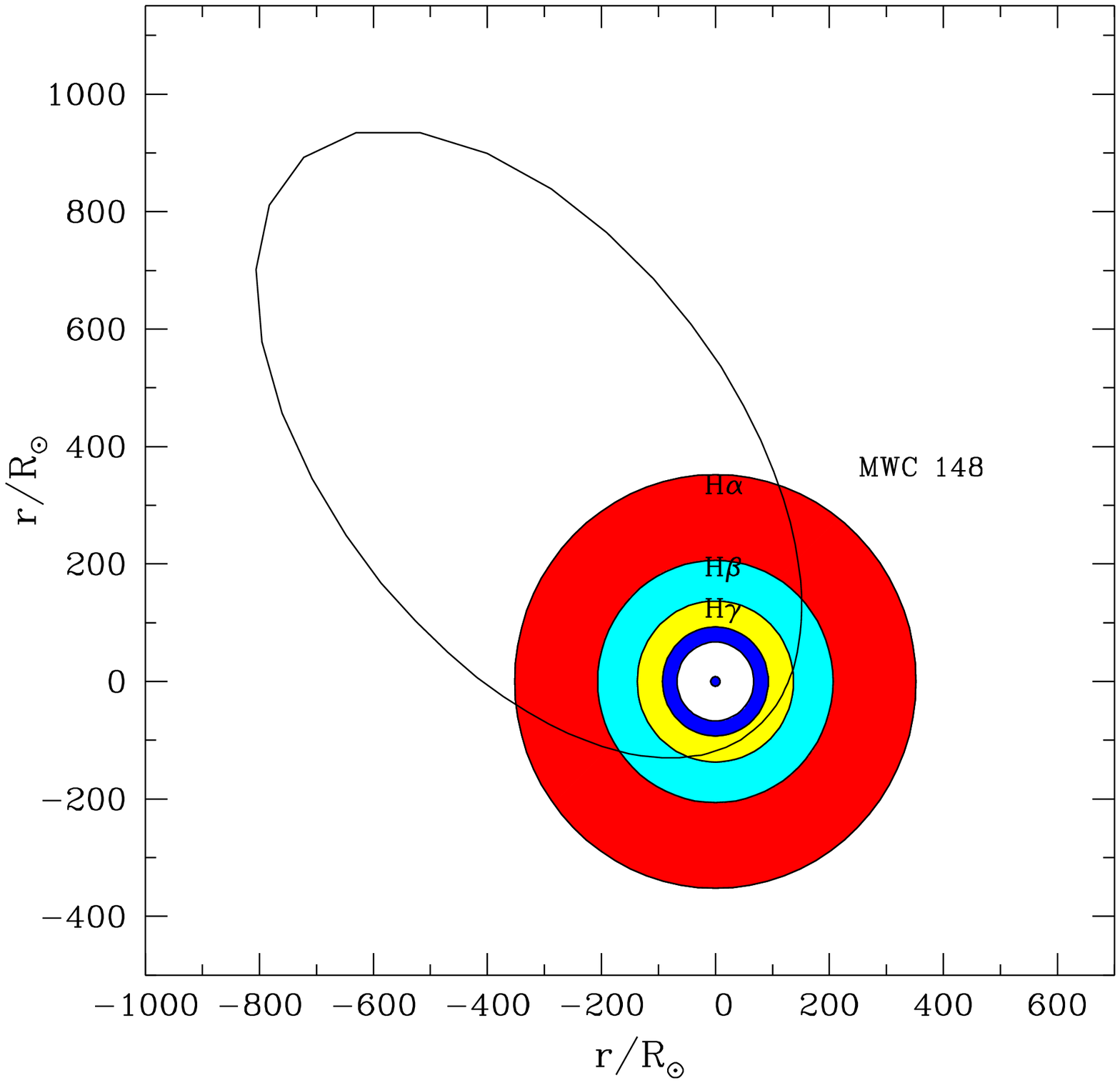}  
   \caption[]{Comparison of the size of the circumstellar disc and the orbit of the
   compact object in the $\gamma$-ray emitting binary MWC 148.  }
   \label{MWC148.orbit}      
 \end{figure*}	     

\section{Relation between  FWHM, EW and  $v\: \sin i$}

Hanuschik (1989) gives the relation for $H\alpha$ emission line:    
\begin{equation}
 \log [{\rm FWHM}({\rm H}\alpha) / (1.23\;  v\: \sin i + 70)] =  - 0.08 \log {\rm EW} + 0.14, \\
\label{FWHM-vsini-EW}
 \end{equation}
where both the FWHM and \vsi\ are  in \kms, EW is in \AA \ (see also   Reid \& Parker 2012). 
For \lsi\ using Coude spectra with   S/N-ratio $\ge 60$,
we measure  FWHM(\Ha) in the range $12.1 - 13.2$~\AA, and estimate  \vsi$=349 \pm 6$~\kms.
This is in agreement with the previous measurement \vsi=360 \kms\ by Hutchings \& Crampton (1981). 
For MWC~148, we measure  FWHM(\Ha) in the range $10.2 - 10.6$~\AA, and estimate  \vsi\ $= 265 \pm 5$~\kms, a value considerably lower than 
\vsi\ $=373$~\kms\  (Casares et al. 2012).

\section{Circumstellar disc size}
\label{Disk.size}

For rotationally dominated profiles, the peak separation can be regarded as a measure of 
the outer radius ($R_{disc}$) of the \Ha\ emitting disc (Huang, 1972),  
 \begin{equation}
      \left( \frac{\Delta V}
                {2\,v\,\sin{i}} \;\right)
       = \;  \left( \frac {R_{disc}}{R_1}\;\right)^{-0.5} ,
  \label{Huang}
  \end{equation}
Eq. \ref{Huang} relies on the assumptions that: (1) the Be star is rotating critically, 
(2) the circumstellar disc is Keplerian, 
and (3) that the line profile shape is dominated by kinematics, and radiative transfer does not play a role. 
$R_{disc}$ can also be calculated as
 \begin{equation}      
      R_{disc} = \frac{ G M_1 \sin ^ 2 {i}} {(0.5 \; \Delta V)^2}, 
  \label{Huang3}
  \end{equation}
where $M_1$ is the mass of the primary, $R_1$ is its radius, and $i$ is the angle of inclination to the line of sight.  
If the parameters ($M_1, R_1, \sin {i}$) are well measured Eqs. \ref{Huang} and \ref{Huang3}
should give similar results. The estimated values of disk radius ($R_{disc} / R_1 $ and  $R_{disc} / R_\odot $) 
are given in Table~\ref{table2}. 

The peak separation in  different lines is (Hanuschik, Kozok \& Kaiser \ 1988):
 \begin{eqnarray}      
   \Delta V ({\rm H}\beta) \approx 1.8 \Delta V ({\rm H}\alpha)                                            \label{H3.1}    \\
   \Delta V ({\rm H}\gamma) \approx 1.2 \Delta V ({\rm H}\beta) \approx 2.2 \Delta V ({\rm H}\alpha)       \label{H3.2}    \\
   \Delta V ({\rm FeII}) \approx 2 \Delta V ({\rm H}\alpha)                                                \label{H3.3}    \\ 
   \Delta V ({\rm FeII})  \approx 1.1  \Delta V ({\rm H}\beta)                                             \label{H3.4}    
 \end{eqnarray}
where Eq. \ref{H3.4} is derived from Eqs. \ref{H3.1} and \ref{H3.2}. 

For MWC~148,  we  calculate  the ratio $\Delta V ({\rm Fe II}) /  \Delta V ({\rm H}\alpha)   \approx  1.96 \pm  0.15 $, a value similar 
to the average value for the Be stars. 
For  MWC~148 and \lsi,  we  also calculate  $\Delta V ({\rm H}\beta) /  \Delta V ({\rm H}\alpha)   \approx  1.35 $,
a value  considerably below the average value for the Be stars.  

For  $\gamma$~Cas and MWC~148,  we  calculate  
$\Delta V ({\rm Fe II}) /  \Delta V ({\rm H}\beta)  \approx  1.5 $,
a value greater than the average value for the Be stars. 
It seems that the relative size of their discs differs from those 
of the classical Be stars, probably due to tidal effects caused by the secondary. 

For the X-ray emission of $\gamma$~Cas two mechanisms have been put forward, accretion of circumstellar disc matter onto an 
orbiting white dwarf, or magnetic field interaction between the star and the circumstellar disc (Smith \& Robinson 1999).
Recently, Motch, Lopes de Oliveira, \& Smith (2015) proposed that in $\gamma$~Cas 
the magnetic field emerges from equatorially condensed subsurface 
convecting layers the thickness of which steeply increases 
with rotation rate. These authors also propose that $\gamma$~Cas and its analogs are
the most massive and closest to critical rotation Be stars.

\section{"Wine bottle"-type emission line}
\label{wine}

In Fig. \ref{fig.e},  it is visible that our observations are with sufficiently high-resolution and
high S/N ratio to detect "wine bottle"-type flank inflections in H$\alpha$ emission of  $\gamma$ Cas 
and MWC~148. There are two possible interpretation of the 
so called "wine bottle" emission-line profiles in Be stars:
{\bf (i)} two component disc model (Kogure 1969), where flank inflections are
interpreted as due to superposition of two concentric disc components separated by a gap;
{\bf (ii)} due to  line radiation transfer effects in an homogeneous, single-component, vertically extended Keplerian
disc as suggested by 
three-dimensional radiative line-transfer calculations (Hummel \& Dachs 1992). 
Symmetric flank inflections of the line profiles are then caused by interaction of non-coherent scattering of line photons in
direction perpendicular to the disc plane with rotational broadening. 

At this stage, we consider that both are possible reasons for the "wine bottle"-type H$\alpha$ emission profile of 
MWC~148. 

Shrader et al. (2015)  pointed out a curious resemblance between the mean 20-60~keV X-ray luminosity 
of $\gamma$ Cas and  \lsi. Here we point a clear similarity between the optical emission lines 
of $\gamma$ Cas and  MWC~148 (see Fig.~\ref{fig.e}). Practically all the emission lines of hydrogen, helium,
FeII  have nearly the same relative intensities and profiles.  This indicates that the 
circumstellar disc in these two stars are very similar -- 
close size, density, mass, temperature,  similar inclination to the line of sight, etc. These 
coincidences render   $\gamma$ Cas  a promising target for future
TeV observations, and perhaps even a test bench for the future understanding of gamma-ray binaries.

Let us assume that "wine bottle"-type flank inflections in H$\alpha$ emission of MWC~148 are due to
an inner disc/ring. Under this hypothesis, we  estimate its radius as $R_{ring} \approx 68$~\rsun\, 
which is close to the distance between the components at periastron passage $r_{per} \approx 86$~\rsun\ (Casares et al. 2012).

\section{The $\gamma$-ray binary MWC 148: disk, inclination, resonances} 
The measured sizes of the circumstellar disc visible in different emission lines in MWC 148
are graphically illustrated in Figs. \ref{fig.MWC148} and \ref{MWC148.orbit} from  edge-on and face-on perspectives, respectively.
Our spectrum is obtained at orbital phase 0.84, which is close to the phase of the periastron $\phi_{per} \approx 0.97$ (Casares et al. 2012).

A part of the BeX$\gamma$B have elliptical orbits and there are evidences that some systems have a misalignment
between the spin axis of the star and the spin axis of the binary orbit. The eccentricities in these
systems are caused by a kick to the neutron star during the supernova event that formed the system. Such kicks
would also give rise to misalignments (Martin, Tout \& Pringel 2009). 
The inclination of the circumstellar disc of MWC~148 to the line of sight is probably similar to that of $\gamma$~Cas ($i \approx 45^0$, Harmanec 2002). 
Assuming $M_1 \approx$ 16~\msun, $M_2 \approx$ 4~\msun\ and $a_1 sin~i \approx 77.6$ ~\rsun\
(see Casares et al. 2012), we estimate the inclination of the orbit $i \sim 47^0$.
It means that in the case of MWC~148, the spin axis of the star approximately coincides with the spin axis of the binary orbit.

For the Be/X-ray binaries, the measurements of disc emission fluxes and disk radii 
show a tendency to have a preferred value (and in the most of cases 
to be around this value, for example see the histogram given in  Fig. 6 by Zamanov et al. 2013), in other words to cluster at specified levels. 
Haigh, Coe \& Fabregat (2004) and Coe et al. (2006) discussed that 
this tendency is related to the presence of resonances between the disc gas and neutron star orbital periods. These resonances
tend to truncate the disc at specific disc radii (Okazaki \& Negueruela 2001). 
The corresponding
values can be calculated using (e.g. Grundstrom et al. 2007):
  \begin{eqnarray} 
    {R_n^{3/2}} = \frac{(G \: M_1)^{1/2}}{2 \: \pi n} \:  P_{orb}, 
  \label{eq.resona}
  \end{eqnarray} 
where $G$ is the gravitational constant and $n$ is the integer number of disc gas rotational periods  
per one orbital period.  
For the system MWC~148, we calculated the following  resonance radii:
$R_8 =124$~\rsun, 
$R_7 =135.6$~\rsun,
$R_6 =150.3$~\rsun,
$R_5 =169.7$~\rsun,
$R_4 =196.9$~\rsun,
$R_3 =238.5$~\rsun,
$R_2 =312.6$~\rsun, and
$R_1 =496.2$~\rsun. 
Using $e = 0.83$, Casares et al. (2012) have calculated the distance between 
the components in periastron and apastron to be r$_{per}$ $\sim$ 86~\rsun\ and r$_{app}$ $\sim$ 935~\rsun\ respectively.
Following the disc sizes, given in Table.\ref{table1}, it seems that the H$\alpha$-emitting disc is 
truncated by  2:1 resonance, H$\beta$-emitting disc - by 
 4:1 and H$\gamma$-emitting disc - by  8:1 resonance.

Recently, Moritani et al. (2015) have found  H$\alpha$, H$\beta$, and H$\gamma$ line profiles 
to exhibit remarkable short-term variability  after the MWC 148 apastron (phases 0.6--0.7), 
whereas they show little variation near the periastron. 
In contrast to the Balmer lines, no profile variability is seen in any FeII emission line. 
In Fig. \ref{MWC148.orbit} we also plot the orbit of the compact object  in addition to the circumstellar disc. 
Here it is visible how,  at periastron passage,
the compact object deepens inside H$\alpha$, H$\beta$, H$\gamma$-disc emitting material, 
but it does not enter the disc domain where the FeII emission originates.
Consequently, we naturally expect that the orbital motion of the compact object will cause 
profile variations in  H$\alpha$, H$\beta$, and H$\gamma$, but  almost no variations in FeII lines. 
This behaviour is partially in agreement with the results of Moritani et al. (2015). 
Future investigations are required to answer whether H$\alpha$, H$\beta$ and H$\gamma$ also vary at periastron. 

\vskip 0.2cm
{\bf Conclusions:}
We report our first observations of Be/X-ray and Be/$\gamma$-ray binaries with the new EsPeRo spectrograph, recently commissioned on the
2.0m RCC telescope at NAO Rozhen, Bulgaria. Our main results at present can be summarized as:
\begin{enumerate}
\item We find a curious similarity between  the optical emission lines of the well known Be star $\gamma$~Cas and  
$\gamma$-ray binary MWC~148, which means that the circumstellar discs in these two objects have similar physical parameters. 
\item For four objects (LSI+61$^{\circ}$303, $\gamma$~Cas, MWC~148, 4U~2206+54),  
we measure emission line parameters and estimate the size of the circumstellar discs. 
\item For MWC~148, our estimates show that at the periastron passage 
the compact objects  deeply penetrates in the circumstellar disc, that can cause changes 
in the Balmer emission line profiles. 
\end{enumerate}
For the purpose of data reduction, we present the identification of a few orders of the comparison spectrum 
of the Thorium-Argon lamp (see Appendix).

 
\vskip 0.2cm 
{\small
{\bf Acknowledgments: }
JM and RZ acknowledge partial support by grant AYA2013-47447-C3-3-P from the Spanish
Ministerio de Econom\'{\i}a y Competi\-tividad (MINECO). JM is a member of the FQM-322 research group
funded by the Consejer\'{\i}a de Econom\'{\i}a, Innovaci\'on, Ciencia y
Empleo of Junta de Andaluc\'{\i}a, and FEDER funds. The manuscript was 
presented as a poster at the 10th conference of Astronomical Society of Bulgaria (Belogradchik, 2-5 July 2015).
}





\vskip 0.2cm

{\bf Appendix: Thorium-Argon spectrum for wavelength calibration}.

 \begin{figure}    
   \vspace{18.5cm}   
     \includegraphics{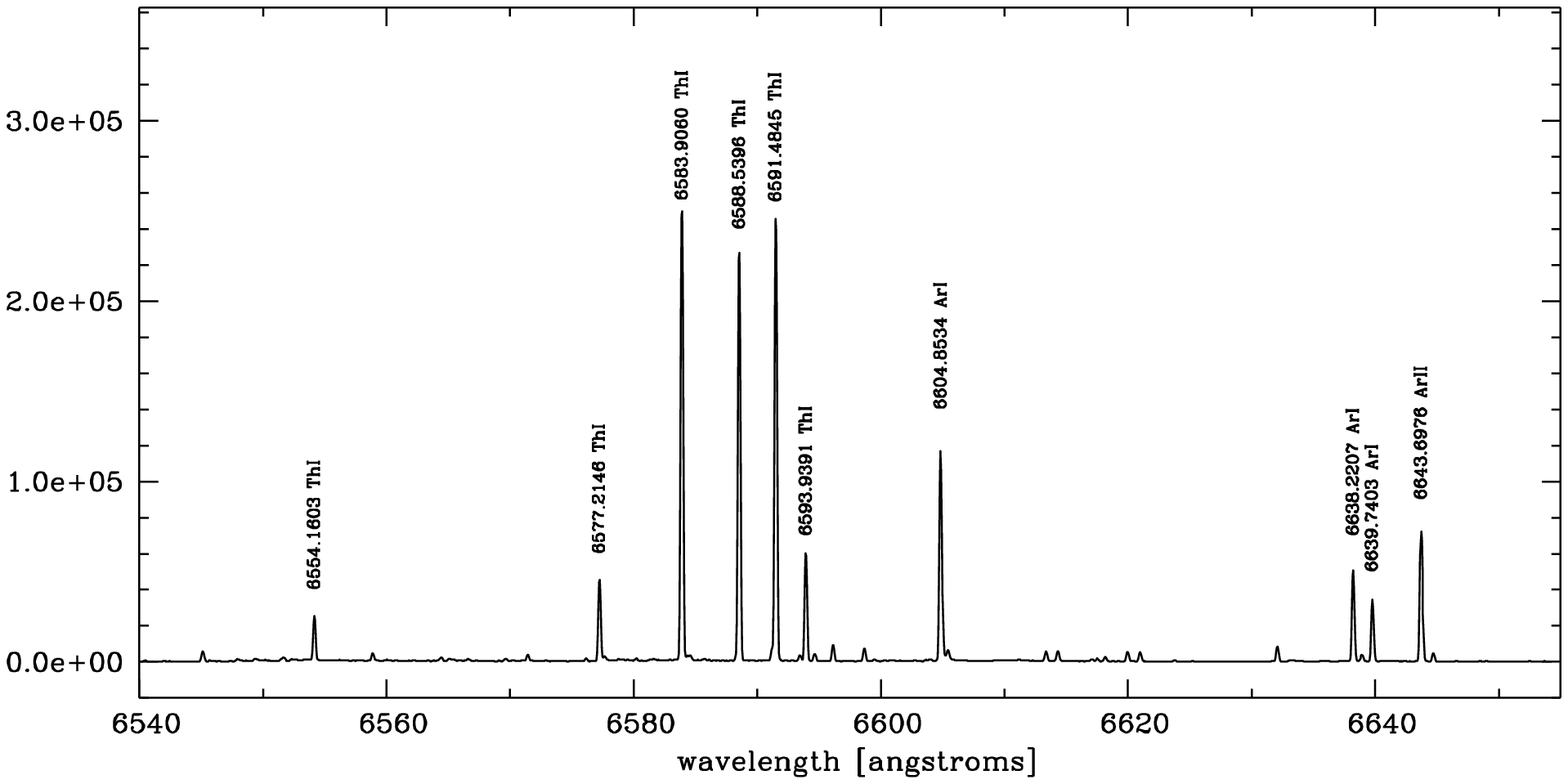}    
     \caption[]{Thorium-Argon Spectrum in the vicinity of H$\alpha$ line ($\lambda 6562$~\AA).}
   \label{ThAr1}      
 \end{figure}	     

 \begin{figure}    
   \vspace{18.5cm}   
     \includegraphics{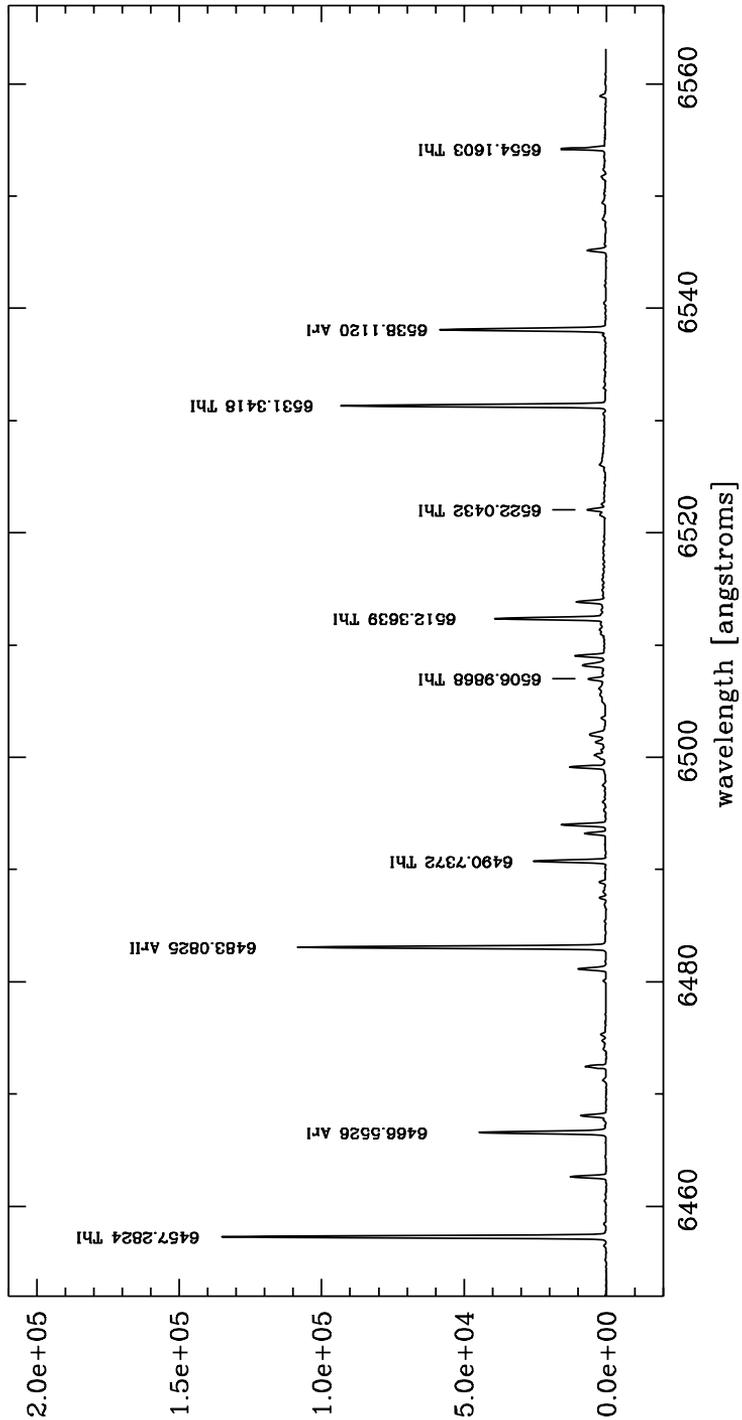}    
     \caption[]{Thorium-Argon Spectrum around $\lambda 6500$~\AA\  \hskip 0.1cm  (order $\# H\alpha +1$) with  some  prominent features identified.}
   \label{ThAr2}      
 \end{figure}	     

 \begin{figure}    
   \vspace{18.5cm}   
     \includegraphics{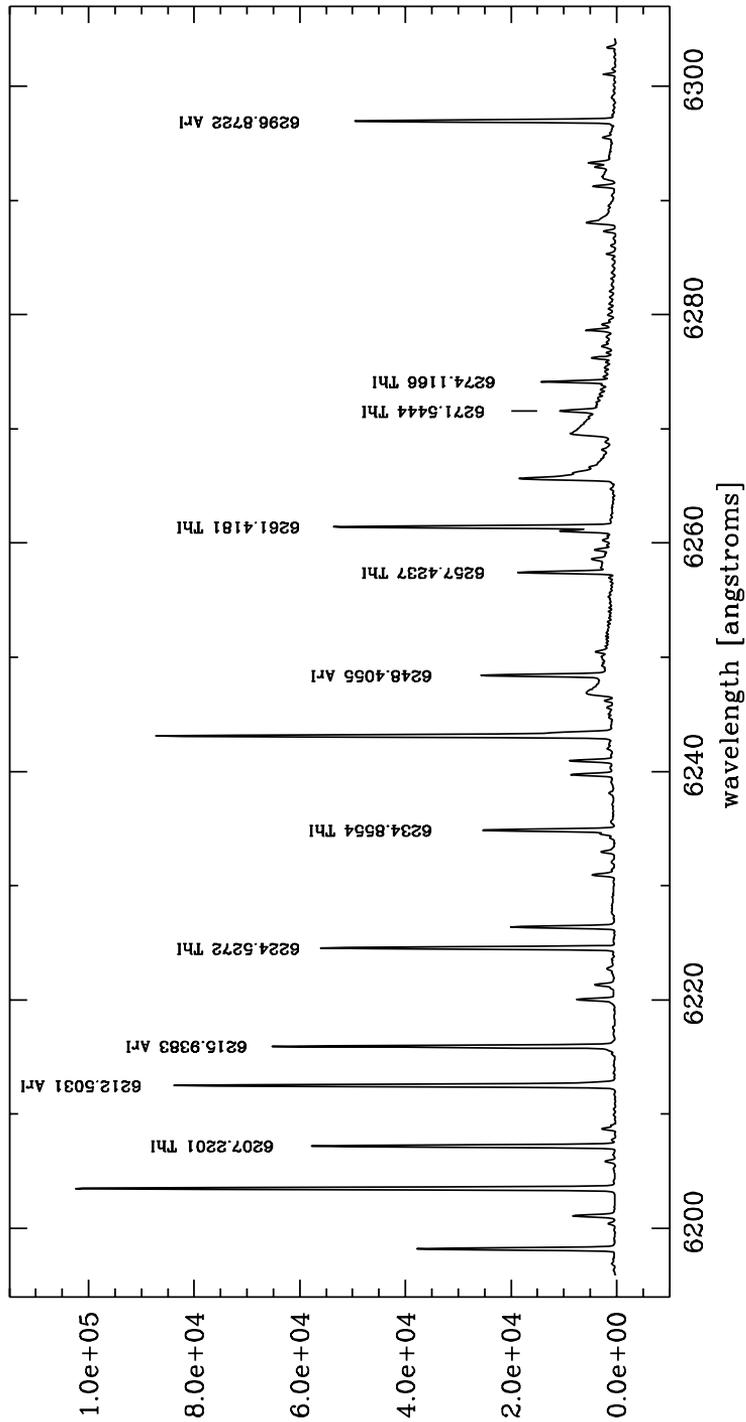}    
     \caption[]{Thorium-Argon Spectrum around $\lambda 6250$~\AA, \hskip 0.1cm order $\# H\alpha +4$. }
   \label{ThAr3}      
 \end{figure}	     

 \begin{figure}    
   \vspace{18.5cm}   
     \includegraphics{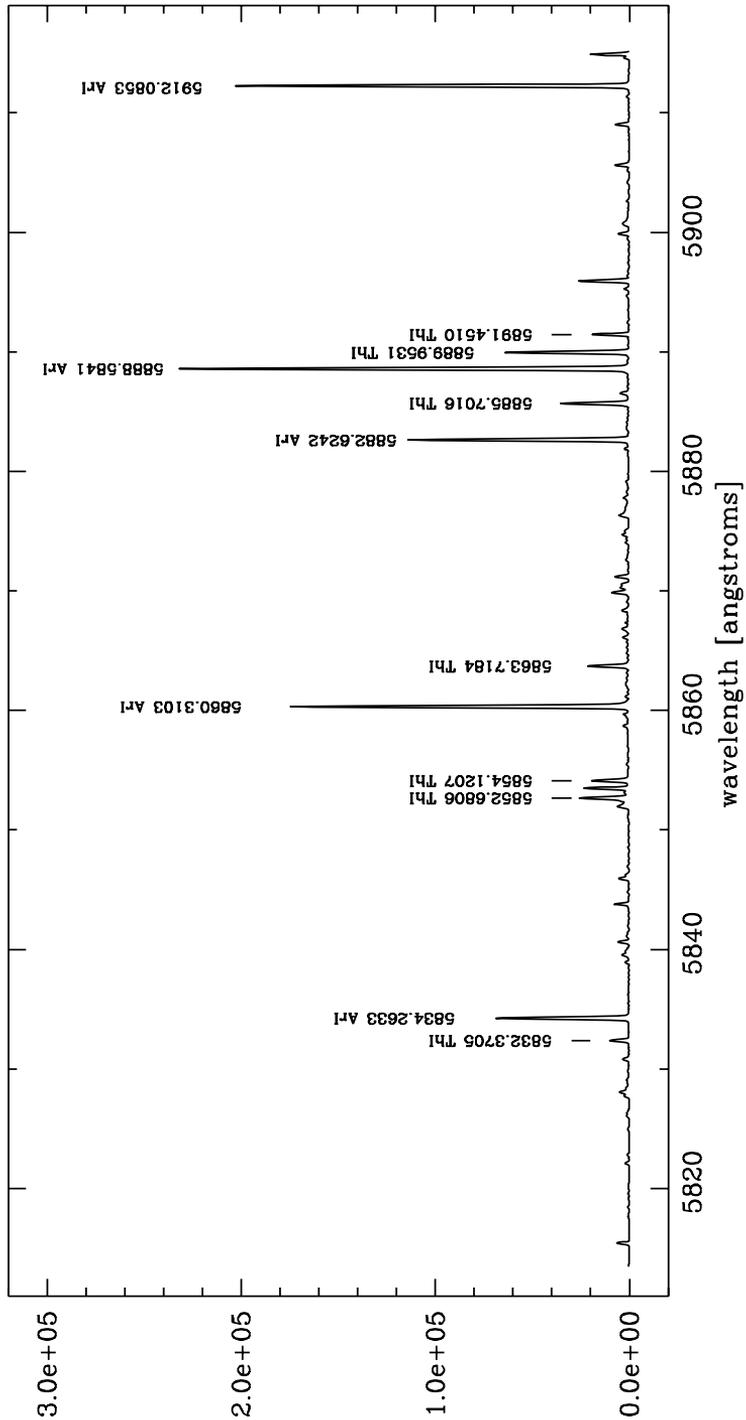}    
     \caption[]{Thorium-Argon Spectrum in the vicinity of NaD ($\lambda 5890$~\AA, \hskip 0.1cm order $\# H\alpha +8$). }
   \label{ThAr4}      
 \end{figure}	     

 \begin{figure}    
   \vspace{18.5cm}   
     \includegraphics{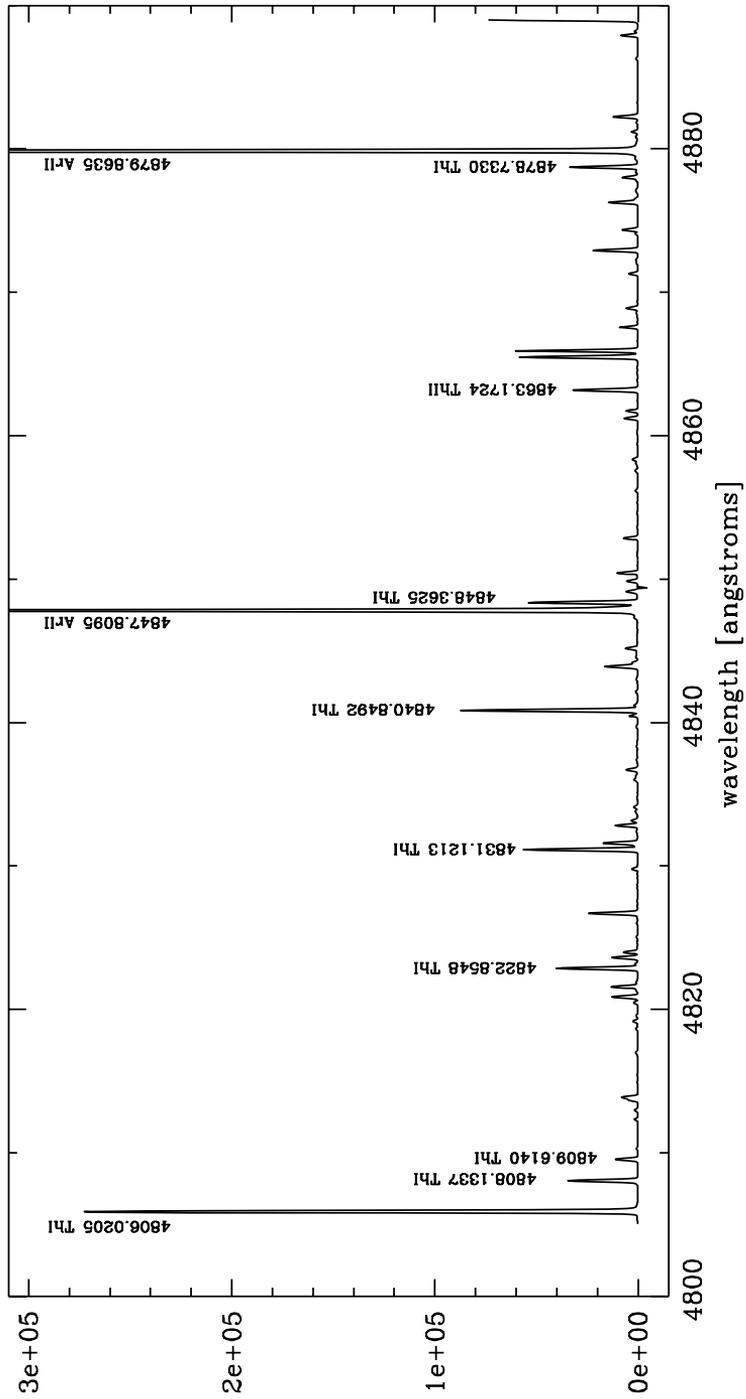}    
     \caption[]{Thorium-Argon Spectrum in the vicinity of H$\beta$ line ($\lambda 4861$~\AA). }
   \label{ThAr5}      
 \end{figure}	     

\end{document}